\documentclass[]{aa}

\usepackage{graphicx}
\usepackage{txfonts}
\usepackage{booktabs}
\usepackage{amsmath}	

\makeatletter
\def\NAT@nmfmt#1{{\NAT@up#1}}
\makeatother

\usepackage[colorlinks=true,allcolors=blue]{hyperref}

\usepackage{acronym}
\newacro{DECam}[DECam]{Dark Energy Camera}
\newacro{igwn}[IGWN]{International Gravitational-Wave Observatory Network}
\newacro{O4}[O4]{observing run 4}
\newacro{GW}[GW]{Gravitational wave}
\newacro{LVK}[LVK]{LIGO-Virgo-KAGRA}
\newacro{DELVE}[DELVE]{DECam Local Volume Exploration survey}
\newacro{des}[DES]{Dark Energy Survey}
\newacro{cbc}[CBC]{compact binary coalescence}
\newacro{em}[EM]{electromagnetic}
\newacro{gwtc5}[GWTC-5.0]{Gravitational Wave Transient Catalog 5.0}
\newacro{ligo}[LIGO]{Laser Interferometer Gravitational-wave Observatory}
\newacro{lsst}[LSST]{Legacy Survey of Space and Time}
\newacro{bbh}[BBH]{binary black hole}

\usepackage{xcolor}
\newcommand{\TODO}[1]{{\color{red}\textbf{[TODO]} #1}}

\makeatletter
\newcommand{\aaDropStarSection}{\@ifstar{\@gobble}{\@gobble}}
\makeatother

\begin{document}


\newcommand{\nkmeans}{20}                 
\newcommand{\mthr}{20}                 
\newcommand{\mthrbright}{15.5}         
\newcommand{\zmin}{0}                  
\newcommand{\zmax}{0.5}                
\newcommand{\mingals}{100}             
\newcommand{\nsideval}{128}            
\newcommand{\delveAreaAllBands}{20\,000} 
\newcommand{\delveAreaFootprint}{15\,859}
\newcommand{\desArea}{4\,923}            
\newcommand{\delveTenSigMedian}{23.85} 
\newcommand{\desTenSig}{24.41}         
\newcommand{\nonDesTenSig}{23.68}      

\newcommand{\NbrCBCTotal}{236}      
\newcommand{\NbrCBCBBH}{\TODO{Update}}      
\newcommand{\NbrCBCnonBBH}{\TODO{Update}}      

\newcommand{\zzero}{0.24}
\newcommand{\phiStar}{1.27}     
\newcommand{\phiStarErr}{0.06}
\newcommand{\MstarSch}{-21.01}               
\newcommand{\MstarSchErr}{0.03}
\newcommand{\alphaSch}{-1.17}
\newcommand{\alphaSchErr}{0.01}
\newcommand{\MminSch}{-23.96}                 
\newcommand{\MmaxSch}{-18.00}                 

\newcommand{\HzDarkBrightSpectralOfourb}{$71.4^{+9.6}_{-7.5}$}
\newcommand{\HzDarkBrightDESStaticOfourb}{$71.4^{+8.8}_{-7.1}$}
\newcommand{\HzDarkBrightDESOfourb}{$70.9^{+8.5}_{-7.1}$}
\newcommand{\HzDarkBrightDELVEOfourb}{$72.4^{+9.6}_{-7.6}$}
\newcommand{\HzDarkBrightDELVEiBandOfourb}{$71.6^{+9.0}_{-7.2}$}
\newcommand{\HzDarkBrightDELVEDeBiased}{$70.6^{+8.7}_{-7.0}$}
\newcommand{\HzDarkBrightDELVEstaticSchecterOfourb}{$72.7^{+9.2}_{-7.5}$}

\newcommand{\HzDarkSpectralOfourb}{$68.0^{+14.7}_{-13.2}$}
\newcommand{\HzDarkDESStaticOfourb}{$69.2^{+12.9}_{-12.2}$}
\newcommand{\HzDarkDESOfourb}{$67.2^{+13.0}_{-12.9}$}
\newcommand{\HzDarkDELVEOfourb}{$70.9^{+14.0}_{-13.5}$}
\newcommand{\HzDarkDELVEiBandOfourb}{$69.2^{+13.4}_{-13.1}$}
\newcommand{\HzDarkDELVEDeBiased}{$67.4^{+13.0}_{-11.2}$}
\newcommand{\HzDarkDELVEstaticSchecterOfourb}{$72.4^{+12.6}_{-11.6}$}

\newcommand{\HzDarkFiducial}{\HzDarkDELVEDeBiased}
\newcommand{\HzDarkBrightFiducial}{\HzDarkBrightDELVEDeBiased}
\newcommand{\HzPlanck}{67.4\pm0.5}
\newcommand{\HzShoes}{73.04\pm1.04}

\title{Constraints on the Cosmic Expansion Rate Using the DECam Local Volume Exploration Survey and the Fifth Gravitational Wave Transient Catalog}

\titlerunning{Dark Sirens with DELVE DR3 and GWTC-5.0}

\author{Sean Patrick MacBride\inst{1}\corrauth{sean.macbride@physik.uzh.ch}
        \and Isaac McMahon\inst{1}
        \and Danny Laghi\inst{1}
        \and David Sánchez-Cid\inst{1}
        \and Felipe Andrade-Oliveira\inst{1}
        \and Marcelle Soares-Santos\inst{1}
        \and Maciej Bilicki\inst{2}
        \and Sarah Ferraiuolo\inst{3,4,5}
        \and Surhud More\inst{6}
        \and Rachel Gray\inst{7}
        \and Gergely Dálya\inst{8}
        }

\authorrunning{S. P. MacBride et al.}

\institute{Physik-Institut, University of Zurich, Winterthurerstrasse 190, 8057 Zurich, Switzerland
        \and Center for Theoretical Physics, Polish Academy of Sciences, 02-668, Warsaw, Poland
        \and Aix Marseille Univ, CNRS/IN2P3, CPPM, Marseille, France
        \and Università di Roma La Sapienza, 00185 Roma, Italy
        \and INFN, Sezione di Roma, I-00185 Roma, Italy
        \and Inter-University Centre for Astronomy and Astrophysics, Pune 411007, India
        \and SUPA, University of Glasgow, Glasgow G12 8QQ, UK
        \and Eötvös University, Budapest 1117, Hungary
        }

\date{Received \today}

\abstract
   {Gravitational-wave standard sirens provide a calibration-free measure of distances, offering important input on the long-standing tension between early- and late-Universe measurements of the Hubble constant ($H_0$).}
   {In the absence of an electromagnetic counterpart, the dark siren method can be employed to statistically correlate a gravitational wave event's localization with a galaxy catalog of potential hosts. We present a measurement of the Hubble constant using the DECam Local Volume Exploration (DELVE) Survey Data Release 3 and the fifth Gravitational-Wave Transient Catalog.} 
   {Applying the dark siren method to a large, photometrically heterogeneous footprint required revised catalog-construction choices: a faint apparent-magnitude threshold ($m_r<\mthr$) to control known photometric-redshift bias, a bright apparent-magnitude threshold ($m_r>\mthrbright$) to exclude saturated sources, implementation of an evolving luminosity function parameterization, and de-biasing survey regions based on survey depth and photometric redshift precision.}
   {With these choices applied to DELVE, we obtain $H_0=\,$\HzDarkFiducial$\,\mathrm{km\,s^{-1}\,Mpc^{-1}}$ from dark sirens alone, and $H_0=\,$\HzDarkBrightFiducial$\,\mathrm{km\,s^{-1}\,Mpc^{-1}}$ when combined with the bright siren GW170817. We test other galaxy catalog configuration choices, and find subdominant influence from photometric band choice and luminosity function parameterization on the final $H_0$ posterior.}
   {Our results demonstrate that extending the dark siren method to wider-area, non-uniform galaxy surveys is feasible and provides additional constraining power on $H_0$ relative to the spectral siren method, opening the door for analyses with forthcoming galaxy catalogs from LSST, Euclid, and Roman.}


\maketitle
\nolinenumbers
\section{Introduction}
\label{sec:introduction}

The Standard model of cosmology, $\Lambda$CDM, provides a precise description of the universe and many of its defining features. As observational cosmology has entered a golden age of high-precision, some parameters have shown discrepancies that warrant extensive study and assessment, including the local rate of accelerated expansion, the Hubble constant ($H_0$). $H_0$ measurements derived from the early Universe --- most precisely, the \textit{Planck} cosmic microwave background analysis, infers $H_0 = \HzPlanck\,\mathrm{km\,s^{-1}\,Mpc^{-1}}$ using the temperature power spectrum, E-mode polarization, and cross correlations in the CMB for $\Lambda$CDM \citep{Planck2020} --- disagreeing at $>5\sigma$ with local Universe distance-ladder measurements, most notably the SH0ES result of $H_0 = \HzShoes\,\mathrm{km\,s^{-1}\,Mpc^{-1}}$ \citep{Riess2022}. Despite extensive scrutiny of each approach, no systematic effect or new physics has been identified to resolve the discrepancy.

\ac{GW} standard sirens offer an independent measurement of distances and, by extension, $H_0$. A \ac{cbc} waveform directly encodes the luminosity distance to the source \citep{Schutz1986,HolzHughes2005}. However, a \ac{GW} \ac{cbc} measurement alone cannot provide the source redshift, and three complementary strategies have been developed to measure cosmological parameters using the \ac{GW} waveform information.

The first, \textit{bright sirens}, requires the detection of an \ac{em} counterpart to a \ac{GW}. The identification of a \ac{GW} merger candidate, and the spectroscopic redshift of its host galaxy, provide the highest $H_0$ precision per \ac{GW} event among the three methods. The binary neutron star merger GW170817 remains the only \ac{GW} event with a confirmed \ac{em} counterpart to date, and provided the first standard-siren measurement of $H_0$ \citep{2017ApJ...848L..12A,Abbott2017Nature,Coulter2017,SoaresSantos2017}.

The second method, \textit{spectral sirens}, exploits redshift-dependent features in the detector-frame mass distribution, such as the neutron-star-black-hole mass gap, peaks and gaps in the mass distributions, and other features to statistically infer the population's source-frame masses, and hence the redshift-luminosity-distance relation \citep{ChernoffFinn1993,Markovic1993,Taylor2012,Farr2019,EzquiagaHolz2021,Mastrogiovanni2021,You2021}.

The third method, the focus of this paper, is \textit{dark sirens}. Dark sirens statistically marginalize over galaxies consistent with a \ac{GW} event's three-dimensional localization volume, weighting each by its probability of hosting the merger \citep{MacLeodHogan2008,Oguri2016,Gair2023,Borghi2024}. This approach requires no \ac{em} counterpart, and applies to a majority of \ac{cbc} detections, with the requirement of sufficiently deep and complete galaxy catalogs over well-localized \ac{GW} volumes.

Dark siren analyses using joint \ac{GW} population and cosmology inference frameworks
have progressed from single-\ac{GW}-event studies (the \ac{des} analyses of GW170814, \citealt{SoaresSantos2019}, and GW190814, \citealt{Palmese2020}), to full-catalog joint analyses across \ac{LVK} observing runs \citep{Abbott2021GWTC2Dark,LVK2025O4aCosmo,GWTC5_cosmo}. Population-level analyses have relied primarily on the all-sky, shallow galaxy catalog data from the GLADE+ compilation \citep{Dalya2018,Dalya2022}. 

These three methods are not strictly independent, as the dark siren method is commonly used in regions where galaxy catalog coverage exists, and the spectral siren method is applied in regions where it does not. This dark and spectral siren constraint can then be combined with bright siren measurements to provide a joint constraint on $H_0$. In addition, recent $H_0$ measurements have shown that standard sirens can be used in combination with other cosmological probes, such as weak lensing, enabling constraints on $\Lambda$CDM parameters beyond $H_0$ \citep{2026A&A...709L..14A}.

\citet{McMahon2026DESGWTC4} demonstrated that a deep, well-characterized photometric galaxy catalog can provide similar constraints to shallow, wide-area dark-siren galaxy catalogs. Combining the \ac{des} Year 6 (Y6) Gold catalog \citep{Bechtol2026DESY6Gold} with GWTC-4.0 \citep{LVK2025GWTC4Props}, \cite{McMahon2026DESGWTC4} demonstrated that DES constrains $H_0$ tighter than GLADE+, despite \ac{des} covering only $\sim\desArea\,\mathrm{deg}^2$ of sky ($\sim12\%$ of the full sky). 

This paper extends the \ac{des} Y6 Gold dark-siren analysis to the \ac{DELVE} Data Release 3 (DR3) catalog \citep{DrlicaWagner2022DELVEDR2,DrlicaWagnerDELVEDR3,Tan2025}, shown in Figure~\ref{fig:delve_coverage}, extending the usable sky area from $\desArea\,\mathrm{deg}^2$ to $\delveAreaFootprint\,\mathrm{deg}^2$, while retaining an identical photometric model, object classifier, and photometric-redshift pipeline as \ac{des} Y6 Gold. This larger footprint covers several \ac{GW} events, increasing the fraction of \ac{GW} events with galaxy catalog information. Because \ac{DELVE} is shallower and photometrically noisier outside the \ac{des} region, extending the analysis to this larger area requires revisiting several of the catalog-construction choices used previously for \ac{des}.

The remainder of this paper is organized as follows. Section~\ref{sec:data} describes the \ac{GW} event catalog and the \ac{DELVE} DR3 galaxy catalog used in this analysis. Section~\ref{sec:methods} presents the dark-siren analysis framework, the construction of the line-of-sight redshift prior, the luminosity-function fit, the $K$-correction procedure, and the \ac{DELVE}-specific catalog choices required to prepare the catalog for dark siren measurement. Section~\ref{sec:results} presents the $H_0$ constraints and their differences under different galaxy catalog analysis choices. Section~\ref{sec:discussion} discusses the implications and limitations of the current analysis. Section~\ref{sec:conclusion} summarizes our results and outlines the path toward deeper and wider catalogs for future dark siren constraints.

\section{Data}
\label{sec:data}

\subsection{The Fifth Gravitational-Wave Transient Catalog}\label{sec:data:gw}

The \ac{LVK} detector network comprises two \ac{ligo} detectors in the United States \citep{Aasi2015}, the Virgo observatory in Italy \citep{Acernese2015}, and the KAGRA observatory in Japan \citep{Akutsu2021}. The most recent release, \ac{gwtc5} \citep{LVKGWTC5}, includes all data taken up to January 28, 2026, bringing the total number of \acp{cbc} to \NbrCBCTotal. In this analysis, the GW event GW170817, a binary neutron-star merger with the only electromagnetic counterpart detection to date, is treated as a bright siren and separately from the other GW merger candidates.

For each GW candidate, we use the parameter estimation posterior samples and \texttt{HEALPix} \citep{Gorski2005} sky localization maps,\footnote{\href{https://zenodo.org/records/20348005}{https://zenodo.org/records/20348005}, \href{https://zenodo.org/records/20348006}{https://zenodo.org/records/20348006}} which jointly encode the detector-frame masses, spins, distances, and sky positions inferred directly from the strain data. Selection effects are modeled using the corresponding set of simulated signal injections, which characterize the detector network sensitivity as a function of source parameters, and are used to compute the selection function described in Section~\ref{sec:methods:framework}. Both signal injections and parameter estimation posterior samples were released as part of \ac{gwtc5} \citep{2026arXiv260527226T}.\footnote{\href{https://zenodo.org/records/19500064}{https://zenodo.org/records/19500064}} For the cosmological inference, we use the FULLPOP-4.0 compact-binary mass population model \citep{LVK2025O4aCosmo} with a Madau-Dickinson merger-rate evolution \citep{MadauDickinson2014}.

\begin{figure*}
    \centering
    \includegraphics[width=\linewidth]{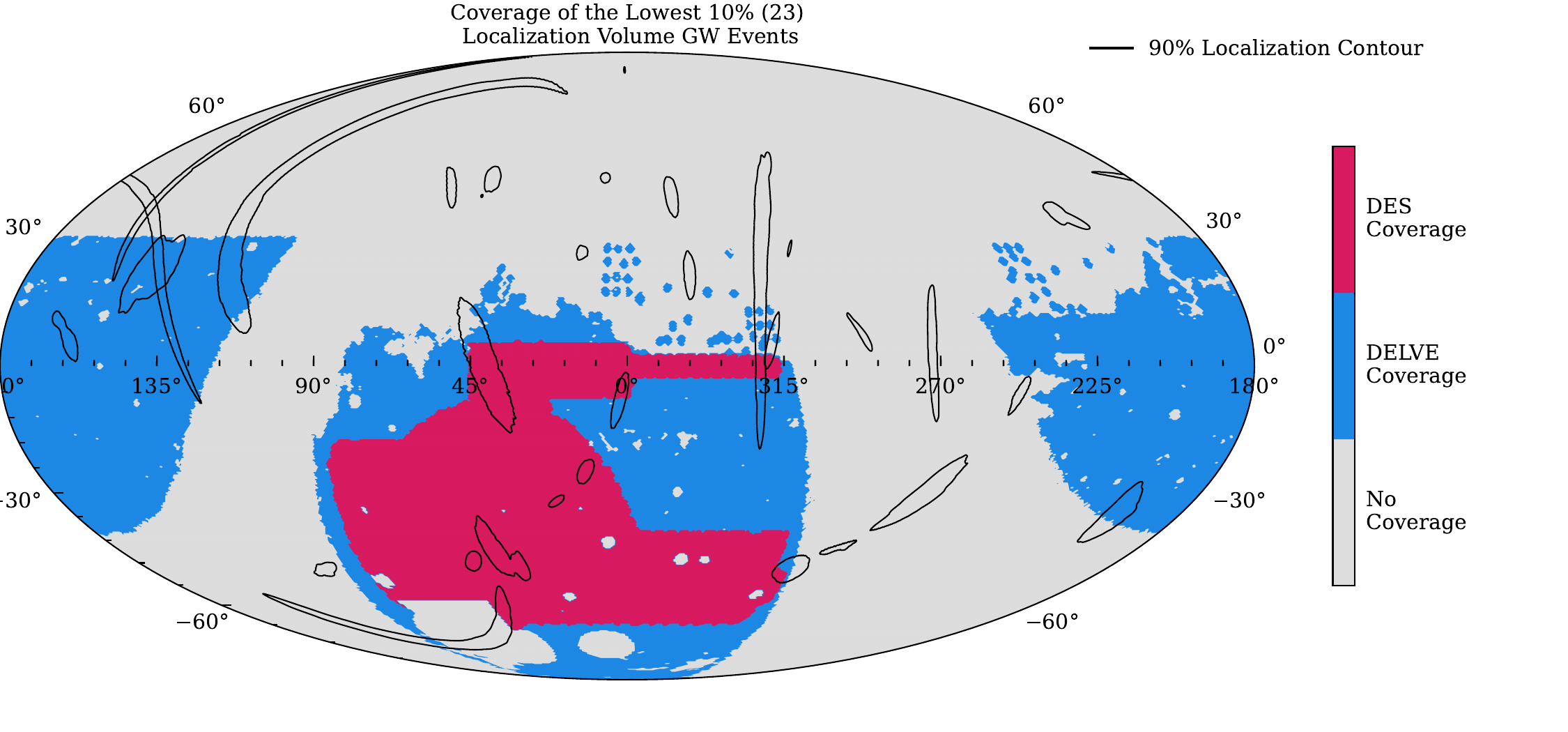}
        \caption{Sky coverage of DELVE DR3 (blue) and DES Y6 Gold (magenta, a subset of the DELVE footprint) in a Mollweide projection, overlaid with the 90\% credible localization contours of the 10\% best-localized events in our GW sample (\ac{gwtc5}). Several of the best-localized events fall outside the \ac{des} footprint, but within \ac{DELVE}.}
    \label{fig:delve_coverage}
\end{figure*}

\subsection{DELVE DR3 Galaxy Catalog}
\label{sec:data:delve}

We use galaxies from the \ac{DELVE} (\citealt{DrlicaWagnerDELVEDR3}, building on the DR2 and DR1 releases described in \citealt{DrlicaWagner2022DELVEDR2,DELVE_DR1}). \ac{DELVE} DR3 combines all science-grade exposures taken with the \ac{DECam} on the National Science Foundation Víctor M. Blanco 4-meter Telescope, including the DES Y6 dataset itself as a subset. \ac{DELVE} DR3 has a footprint of $\delveAreaFootprint\,\mathrm{deg}^2$ with $griz$ photometry across most of the southern sky outside the Galactic plane (to declination $\delta \lesssim 30^\circ$), compared to the $\desArea\,\mathrm{deg}^2$ DES footprint (Figure \ref{fig:delve_coverage}).

\ac{DELVE} DR3 has identical data processing to DES Y6 Gold, and we adopt an identical photometric model, object catalog columns, and quality cuts as \citet{McMahon2026DESGWTC4}, namely:
\begin{itemize}
    \item \texttt{BDF\_MAG\_*\_CORRECTED} magnitudes (bulge$+$disc fit photometry, extinction- and zero-point-corrected) in $g,r,i,z$;
    \item finite (non-\texttt{NaN}) photometry in all four $griz$ bands and a finite and positive photometric redshift;
    \item \texttt{FLAGS\_FOOTPRINT==1}, \texttt{FLAGS\_FOREGROUND==0}, and \texttt{FLAGS\_GOLD==0}, which enforce a minimum exposure depth, exclude regions affected by bright foreground sources, and exclude objects and regions that were the result of data processing failures;
    \item \texttt{EXT\_XGB==4}, the gradient-boosted decision-tree star/galaxy classifier \citep{ChenGuestrin2016} used in preference to the default morphological classifier for its superior performance to select for an ultra-pure galaxy sample.
\end{itemize}

\begin{figure}
    \centering
    \includegraphics[width=\columnwidth]{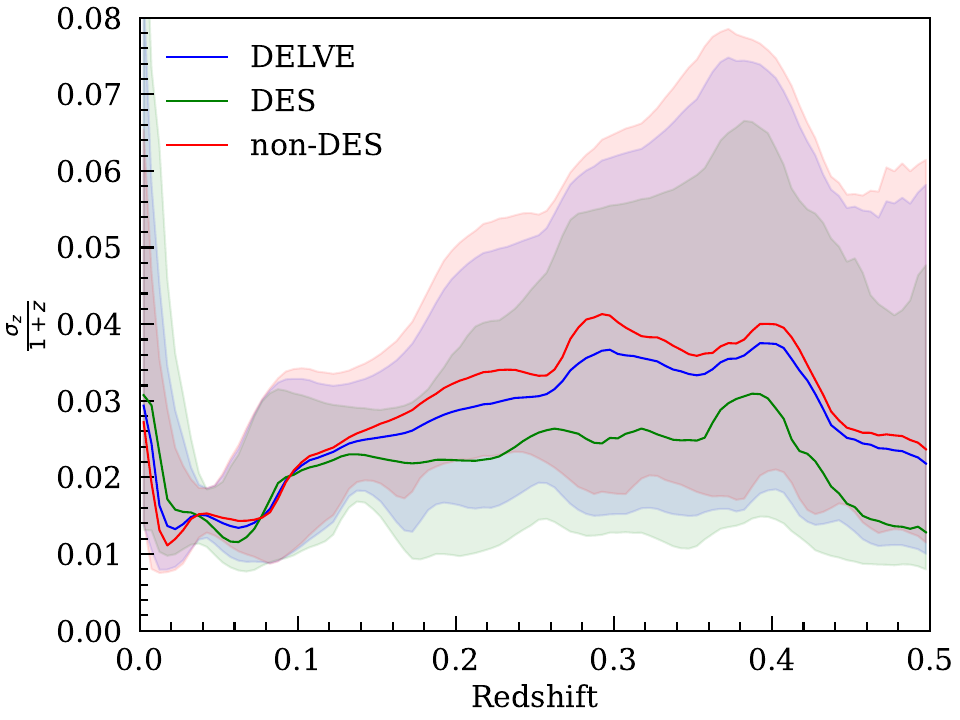}
    \caption{Photometric-redshift scatter $\frac{\sigma_z}{1+z}$ as a function of redshift for galaxies satisfying the galaxy catalog cuts in Section \ref{sec:data:delve}, and $m_r < \mthr$, shown separated by the \ac{des} region, the non-\ac{des} region, and all of \ac{DELVE}. The solid line shows the median $\sigma_z$ at fixed $z$ and the shaded band the $1\sigma$ scatter at that $z$. The non-DES region is photometrically noisier than the DES region for $z \gtrsim 0.1$.}
    \label{fig:photz_uncertainty}
\end{figure}

Photometric redshifts are computed with the same directional-neighborhood-fitting (DNF) algorithm \citep{DeVicente2016} used by DES Y6 Gold. Within the DES footprint, \ac{DELVE} photometry and photometric redshifts are by construction identical to DES Y6 Gold; galaxies outside the DES footprint are processed through the same pipeline but reach a shallower median $10\sigma$ $r$-band limiting magnitude\footnote{$\nonDesTenSig$ mag in the non-DES region versus $\desTenSig$ mag in the DES region; DELVE-wide median $\delveTenSigMedian$ mag}, resulting in a larger photometric-redshift scatter in the non-\ac{des} region, particularly at $z\gtrsim0.1$ (Figure \ref{fig:photz_uncertainty}).



\section{Methodology}
\label{sec:methods}

\subsection{Parameter Inference}
\label{sec:methods:framework}

We follow the joint cosmology and GW population Bayesian inference framework implemented in \texttt{gwcosmo} \citep{Gray2020,Gray2022,Gray2023,Papadopoulos2026}. Given $N_{\rm det}$ detected CBC events with data $\{d\}$, the joint posterior on the CBC population parameters $\Lambda$ (e.g.\ the binary component masses, and redshift-evolving merger-rate distributions) and the cosmological parameters $\Lambda_c$
is

\begin{equation}
\begin{split}
    p(\Lambda, \Lambda_c \mid \{d\}, N_{\rm det}) \propto \pi(\Lambda)\,\pi(\Lambda_c)\,\xi(\Lambda,\Lambda_c)^{-N_{\rm det}} \\ \times \prod_{i=1}^{N_{\rm det}} \int {\rm d}\theta_i^{\rm det}\, \frac{p(\theta_i^{\rm det}\mid d_i)}{\pi(\theta_i^{\rm det})} \left| \frac{{\rm d}\theta_i^{\rm det}(\theta_i^{\rm src},\Lambda_c)}{{\rm d}\theta_i^{\rm src}} \right|^{-1} p(\theta_i^{\rm src}\mid \Lambda),
\end{split}
    \label{eq:joint_posterior}
\end{equation}
where $\pi(\cdot)$ denotes a prior distribution, $\theta_i^{\rm det}$ are the detector-frame parameters recovered directly from the GW strain data for event $i$, $\theta_i^{\rm src}$ are the corresponding source-frame parameters (related to $\theta_i^{\rm det}$ through cosmological parameters $\Lambda_c$), $p(\theta_i^{\rm det}\mid d_i)$ is the single-event likelihood (evaluated via the released posterior samples), and $p(\theta_i^{\rm src}\mid\Lambda)$ is the assumed population distribution of source-frame parameters. The selection function $\xi(\Lambda,\Lambda_c)$ corrects for the detector network's sensitivity as a function of source parameters, and is evaluated using the simulated injections described in Section~\ref{sec:data:gw}.

The redshift dependence of $\theta_i^{\rm src}$ on $\Lambda_c$ is what enables us to constrain $H_0$: at fixed detector-frame chirp mass and luminosity distance, different values of $H_0$ imply different source-frame masses, and the astrophysical population distribution $p(\theta^{\rm src}\mid\Lambda)$ therefore prefers some values of $H_0$ over others even without any galaxy-catalog information. The galaxy catalog information enters through the $p(\theta^{\rm src}_i\mid\Lambda)$ term.


For each \texttt{HEALPix} pixel (\texttt{nside=\nsideval}) along a given GW event's line of sight, the redshift component of the source-frame population prior is constructed as the sum of an in-catalog and an out-of-catalog term,
\begin{equation}
    p(\theta^{\rm src}_i \mid \Lambda) \; \propto \; \frac{{\rm d}N_{\rm gal,cat}^{\rm eff}}{{\rm d}z\,{\rm d}\Omega} \; + \; \frac{{\rm d}N_{\rm gal,out}^{\rm eff}}{{\rm d}z\,{\rm d}\Omega}.
    \label{eq:los_prior},
\end{equation}
respectively. 

The in-catalog term 
sums over galaxies with a measured photometric redshift in the pixel, each represented as a Gaussian likelihood in redshift centered on its photometric-redshift point estimate $z_i$ with width $\sigma_{z,i}$, and weighted by
\begin{equation}
    w_i(\epsilon, M_i) = 10^{-0.4\,\epsilon\,(M_i - M^\ast)},
    \label{eq:lum_weight}
\end{equation}
where $M^\ast$ is the characteristic absolute magnitude of the luminosity function (Section~\ref{sec:methods:lf}) and $\epsilon$ differentiates between weighting each galaxy uniformly ($\epsilon=0$), or by its luminosity ($\epsilon=1$). The convention of luminosity weighting is motivated by the assumption that the compact-binary merger rate traces stellar mass, and hence luminosity. We adopt $\epsilon=1$ for this analysis, and do not consider the $\epsilon=0$ case here, as previous studies have shown that varying $\epsilon$ has a negligible effect on the $H_0$ posterior relative to the current \ac{GW} statistical uncertainties~\citep{GWTC5_cosmo,McMahon2026DESGWTC4}.

The out-of-catalog term 
represents galaxies too faint, too bright, or otherwise excluded from the in-catalog component, modeled as an isotropic population, uniform in comoving volume between the effective survey magnitude limits and the assumed absolute-magnitude range of the luminosity function. 

We only compute the in-catalog term for pixels which have more than 50 galaxies ($\sim2\sigma$ less than the average number of galaxies per pixel), which is a conservative choice to exclude pixels at the boundaries that are partially covered by the survey footprint and include pixels that have a low number of galaxies over the same spatial area as other well-sampled pixels. Pixels without catalog contribution are modeled with the out-of-catalog term alone. 

\subsection{Luminosity function}\label{sec:methods:lf}

We model the galaxy luminosity function following \citet{Schechter1976},
\begin{equation}
\begin{split}
    \Phi(M)\,{\rm d}M = 0.4\ln(10)\,\phi^\ast\,10^{0.4(\alpha+1)(M^\ast-M)}\\
    \times\exp\!\left[-10^{0.4(M^\ast-M)}\right]{\rm d}M,
\end{split}
    \label{eq:schechter}
\end{equation}
with characteristic density $\phi^\ast$, characteristic magnitude $M^\ast$, and faint-end slope $\alpha$. We fit Equation~\ref{eq:schechter} to the DELVE $r$-band photometry using a modified $1/V_{\rm max}$ estimator derived from \citet{Eales1993},
\begin{equation}
    \int_{M_1}^{M_2} \Phi(M)\,{\rm d}M = \sum_{i}^{N} \frac{P_i}{C_{s,i}\,V_{{\rm max},i}},
    \label{eq:vmax}
\end{equation}
summing over the $N$ galaxies in each absolute-magnitude bin $[M_1,M_2]$. The changes to the \citet{Eales1993} $1/V_{\rm max}$ estimator are carried over from \citet{McMahon2026DESGWTC4}: firstly, we represent each galaxy's redshift as a Gaussian probability density $\mathcal{N}(z,z_i,\sigma_{z,i})$ rather than a point estimate, so that the probability of a galaxy lying within a given redshift bin $[z_1,z_2]$ is $P_i = \int_{z_1}^{z_2}\mathcal{N}(z,z_i,\sigma_{z,i})\,{\rm d}z$; secondly, we include a $C_{s,i}$ term, which is a magnitude- and surface-brightness-dependent completeness correction \citep{Capozzi2017}, computed from the Balrog Y6 synthetic source injection catalog from \ac{des}, which characterizes the object detection rate as a function of apparent magnitude and surface brightness \citep{Anbajagane2025}. 

The maximum accessible comoving volume for each galaxy,
\begin{equation}
    V_{{\rm max},i} = A_{\rm cat}\left[\frac{c}{H_0}\int_{z_{\min,i}}^{z_{\max,i}} \frac{{\rm d}z'}{\sqrt{\Omega_m(1+z')^3 + (1-\Omega_m)}}\right]^3,
    \label{eq:vmaxi}
\end{equation}
is computed over the solid angle $A_{\rm cat}$ of the catalog used for the fit, with $z_{\min,i}=\max(z_1,z_{{\rm bright},i})$ and $z_{\max,i}=\min(z_2,z_{{\rm faint},i})$ set by the redshifts at which the galaxy would exceed the bright and faint apparent-magnitude limits. We adopt $H_0=100\,h\,{\rm km\,s^{-1}\,Mpc^{-1}}$ and matter density parameter $\Omega_m=0.3065$ \citep{Planck2020}.

We fit the luminosity function to the DES footprint only, where the surface-brightness completeness correction is available and photometric-redshift quality is highest, and apply the luminosity function parameters across the full \ac{DELVE} footprint under the assumption that the underlying luminosity function is isotropic. We estimate the luminosity function fit uncertainty by dividing the \ac{des} footprint into \nkmeans\ subregions using a $k$-means algorithm \citep{Suchyta2016,Kwan2017} and computing the mean and scatter of the fit across the subregions.

We also evaluate an evolving ßluminosity function parameterization,
\begin{align}
    M^\ast(z) &= M_0^\ast + Q_M\,(z-z_0), \label{eq:mstar_evolving}\\
    \alpha(z) &= \alpha_0 + Q_\alpha\,(z-z_0), \label{eq:alpha_evolving}\\
    \phi^\ast(z) &= \phi_0^\ast\,10^{P_\phi(z-z_0)},\label{eq:phistar_evolving}
\end{align}
where luminosity function parameters $Q_M$, $Q_\alpha$, and $P_\phi$ evolve linearly with redshift, relative to a reference redshift, $z_0$, and the luminosity function parameters $M^*_0$, $\alpha_0$, and $\phi^*_0$ measured at that reference redshift $z_0$.

Absolute magnitudes are computed from the observed apparent magnitude $m_r$ and luminosity distance $D_L$ via
\begin{equation}
    M_r = m_r - 5\log_{10}\!\left(\frac{D_L}{10\,{\rm pc}}\right) - K(z),
    \label{eq:absmag}
\end{equation}
with
\begin{equation}
    D_L = \frac{c(1+z)}{H_0}\int_0^z \frac{{\rm d}z'}{\sqrt{\Omega_m(1+z')^3+(1-\Omega_m)}}.
    \label{eq:dl}
\end{equation}
We use a $K$-correction parameterization for $K(z)$ from \citet{Chilingarian2010} (hereafter Chilingarian), which is a polynomial fit in redshift and color, applied after transforming \ac{DECam} photometry into the SDSS photometric system \citep{Abbott2021DECamPhot}. The fit is calibrated and validated only over $z<0.5$ and $-0.1 < g-r < 1.9$, which is the origin of the $z_{\max}=\zmax$ limit adopted throughout this work (Section~\ref{sec:methods:catchoices}); this redshift cut removes only a modest fraction of the cumulative GW localization volume overlapping the DELVE footprint.


Our luminosity function parameters are reported in Table~\ref{tab:lf_params} using the median redshift of \ac{DELVE} $z_0=0.24$ for the static luminosity function parameterization, and using $z=0.1$ as our redshift reference for the evolving luminosity function parameterization.

\begin{table*}
    \centering
    \begin{tabular}{c  c c c c c c c c c}
    
    \hline
    \hline
Photometric & $z_0$ & $\phi^\ast$ & $M^\ast - 5\log_{10}h$ & $\alpha$ & $M_{\min}$ & $M_{\max}$ & $Q_M$ & $Q_\alpha$ & $P_\phi$ \\
Band &   & $10^{-2}\ h^3\  {\rm Mpc}^{-2}$ & ${\rm Mag}$ &  & {\rm Mag} & {\rm Mag} &  &  & \\
        \midrule
r & 0.10  & $1.28 \pm\ 0.09$ & $-20.89 \pm\ 0.03$ & $-1.23 \pm\ 0.02$ & $-21.76$ & $-18.00$ & $-0.80 \pm\ 0.03$ & $0.43 \pm\ 0.01$ & $-0.03 \pm\ 0.02$ \\ \hline
r & 0.24 & $1.27 \pm\ 0.05$ & $-21.01 \pm\ 0.02$ & $-1.17 \pm\ 0.01$ & $-23.96$ & $-18.00$ & -- & -- & --\\ \hline
i & 0.10  & $1.07 \pm\ 0.03$ & $-21.33 \pm\ 0.04$ & $-1.16 \pm\ 0.02$ & $-21.72$ & $-18.00$ & $-1.53 \pm\ 0.06$ & $-0.33 \pm\ 0.01$ & $-0.56 \pm\ 0.03$ \\ 
        \bottomrule
    \end{tabular}%
    \caption{Luminosity function parameters and linear evolution parameters for the different photometric band and $K$-correction parameterizations tested. In the static case, luminosity function parameters are evaluated at the median redshift of the DELVE sample, $z_0=0.24$. In the evolving luminosity function case, the luminosity functions are evaluated at $z_0=0.1$ and parameterized linearly in redshift following $Q_M$, $Q_\alpha$, and $P_\phi$.}
    \label{tab:lf_params}
\end{table*}
These values are broadly consistent with published $r$-band determinations \citep{Blanton2003,MonteroDortaPrada2009,Hill2010,Loveday2012,Moore2025,McMahon2026DESGWTC4}.

\subsection{Analysis Choices on the DELVE Galaxy Catalog}\label{sec:methods:catchoices}

We refine three galaxy catalog-construction choices, driven by the fact that the non-\ac{des} region of \ac{DELVE} is on average shallower and photometrically noisier than the \ac{des} region (Section~\ref{sec:data:delve}).

\subsubsection{Faint-end Apparent-Magnitude Threshold} 
\label{sec:methods:catchoices:absmag}
Photometric redshifts can exhibit a bias for a variety of reasons, including a lack of spectroscopic training galaxies, and also because spectral features cross between photometric bandpasses and require additional photometry to break the color degeneracy. 
We impose a requirement of 10\% or less on the mean absolute photometric-redshift bias in redshift-magnitude space, and measure the mean absolute redshift bias by crossmatching the photometric redshift measurements from \ac{DELVE} to an amalgamation of spectroscopic redshift catalogs (SDSS DR17; \citet{SDSS_DR17} OzDES; \citet{ozdes}, PRIMUS; \citet{PRIMUSI}, VIPERS; \citet{VIMOS}, WiggleZ; \citet{WiggleZ}, zCOSMOS; \citet{zCosmos}, VLT/VIMOS; \citet{goods_vmos}, 3D-HST; \citet{3d_hst}, DEEP2; \citet{deep2}, 2dF; \citet{2df}, 6dF; \citet{6df}).
Driven by this requirement, we impose $m_{\rm thr,faint}=\mthr$ in $r$-band, similar to \citet{McMahon2026DESGWTC4}, and $m_{\rm thr,faint}=19.5$ in i-band, such that galaxies fainter than $m_{\rm thr,faint}$ are excluded from the analysis and treated via the out-of-catalog term of Equation~\ref{eq:los_prior} rather than the in-catalog term. This reduces the completeness of \ac{DELVE} relative to using its full depth, but allows the in-catalog redshift range to be safely extended to $z=\zmax$ (the limit otherwise imposed by the $K$-correction validity range of Section~\ref{sec:methods:lf}). 

\begin{figure}
    \centering
    \includegraphics[width=\columnwidth]{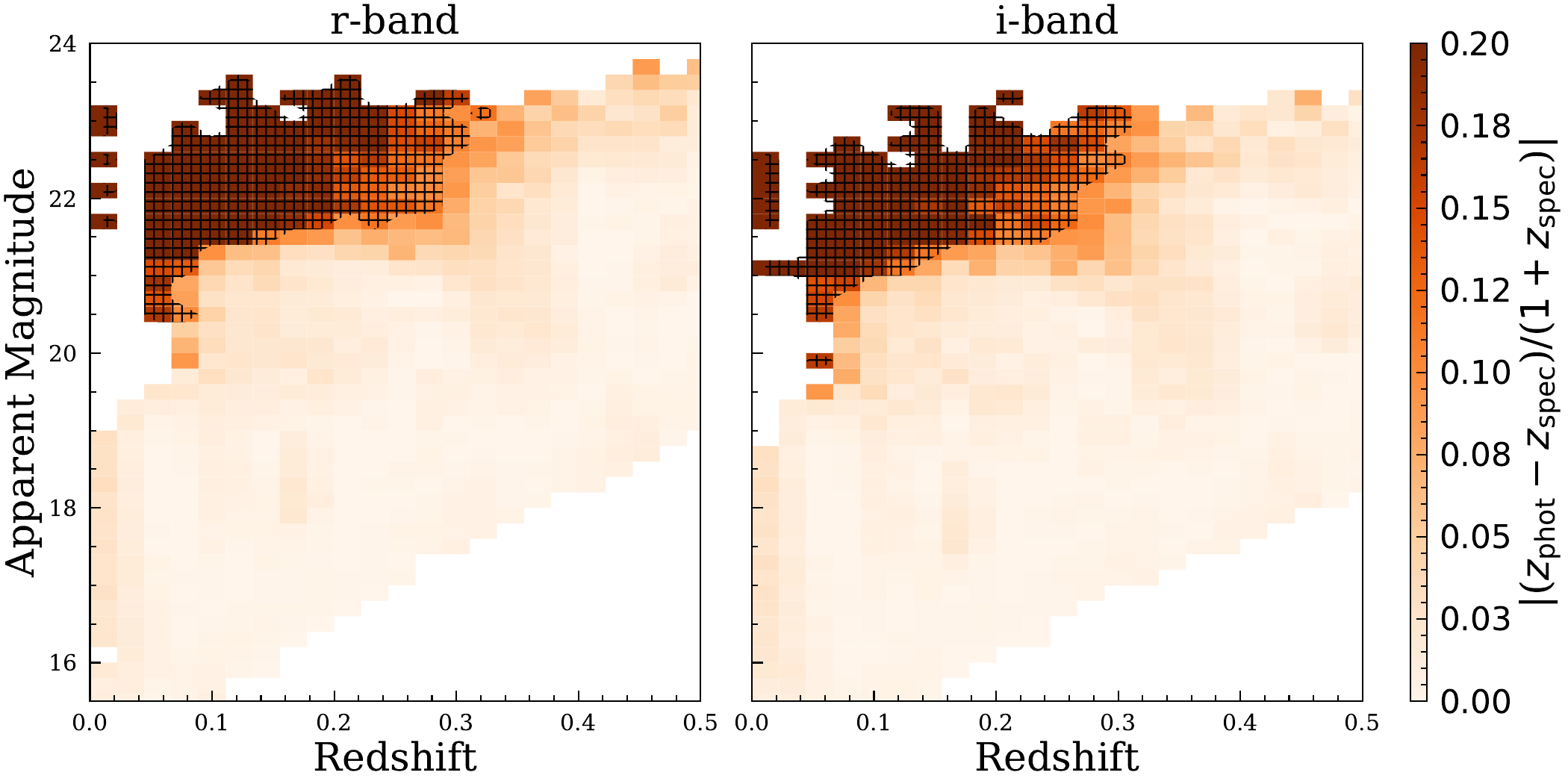}
    \caption{Redshift bias as a function of redshift and apparent magnitude for \ac{DELVE}, across the two photometric bands used in this analysis. The hatched area in each subplot shows the regions where the mean absolute redshift bias is > 10\%.}
    \label{fig:pz_bias_color}
\end{figure}

\subsubsection{Bright-end Apparent-Magnitude Threshold} 

\begin{figure}
    \centering
    \includegraphics[width=\columnwidth]{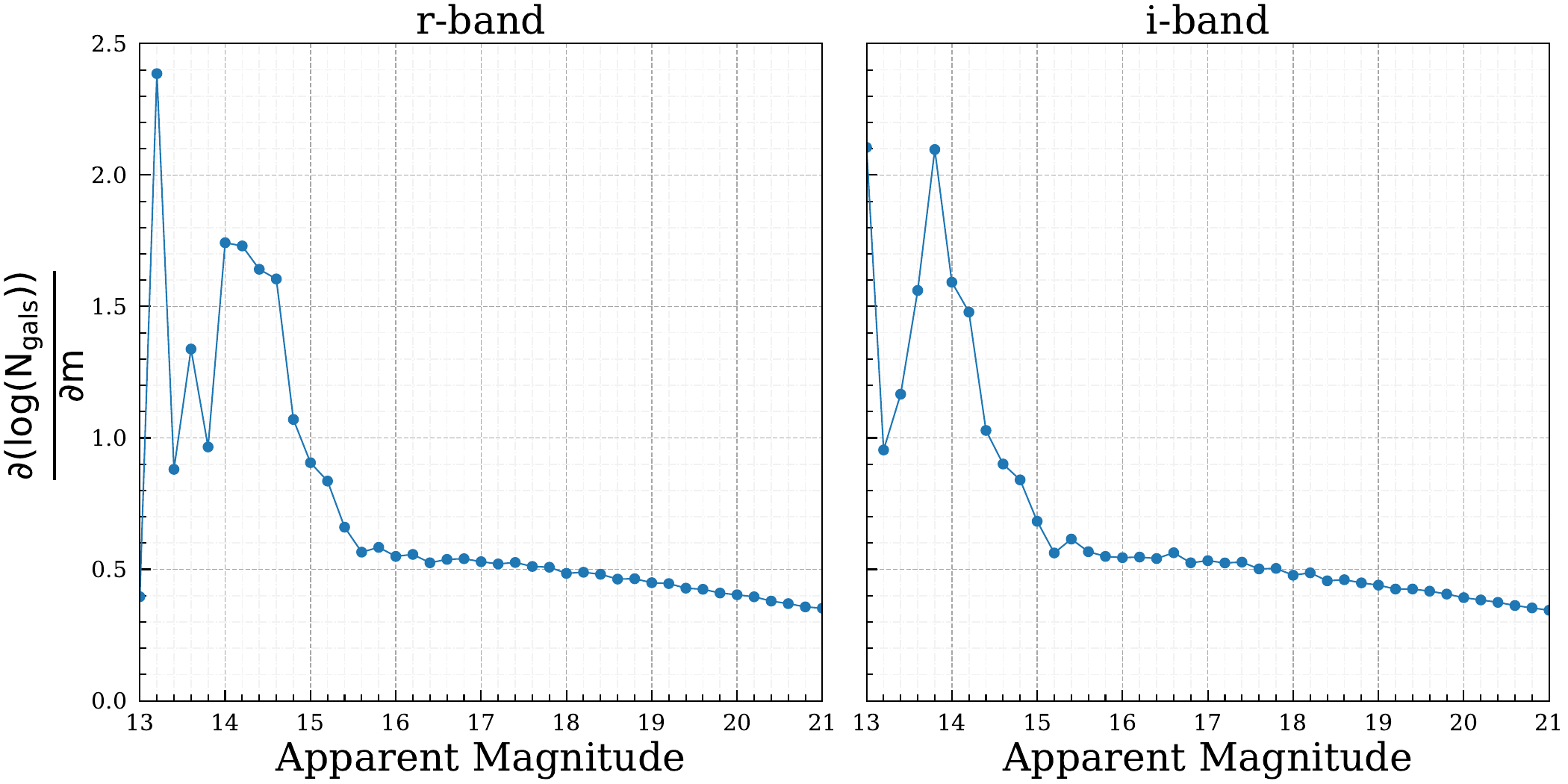}
    \caption{The derivative in magnitude space of DELVE $r$- and $i$-band galaxy number counts, as a function of apparent magnitude. The bright-end threshold is set at the magnitude where $\mathrm{d}\log(N_{\rm gal})/\mathrm{d}m$ starts to steadily decrease due to CCD saturation no longer affecting object identification.}
    \label{fig:mag_cuts}
\end{figure}

Previously, saturation of the \ac{DECam} CCDs at bright magnitudes was handled with a lower redshift cut, $z_{\rm low}=0.05$, since saturation only affects the lowest-redshift, brightest galaxies in a magnitude-limited survey of fixed depth. 
Here, we impose a bright-end apparent-magnitude threshold to exclude bright galaxies that saturate the \ac{DECam} CCDs. 
To define this threshold, we use the smoothness of the derivative of galaxy number counts as a function of magnitude as an indicator of saturation, shown in Figure~\ref{fig:mag_cuts}. 
With this criterion, we select $m_{\rm thr,bright}=\mthrbright$ in $r$-band, and $m_{\rm thr,bright}=15$ for i-band. 
This bright-end apparent magnitude threshold works analogously to the faint-end magnitude threshold term, modifying the out-of-catalog term by considering galaxies brighter than saturation as out-of-catalog. 
After conversion to absolute magnitudes, the bright apparent magnitude threshold excludes all galaxy catalog information below $z\sim0.018$ for $r$-band, and $z\sim0.015$ for $i$-band, limiting the line-of-sight contribution to the out-of-catalog term only, inside this volume.
Inclusion of $z<0.05$ galaxies that do not saturate the DECam CCDs increases the contribution of galaxies in the local universe, crucial for constraining $H_0$ by incorporating galaxy catalog information for well-localized, low-redshift \ac{GW} events.


\subsubsection{Redshift Bias Across Survey Regions} 
\label{sec:methods:catchoices:debias}

\begin{figure}
    \centering
    \includegraphics[width=\columnwidth]{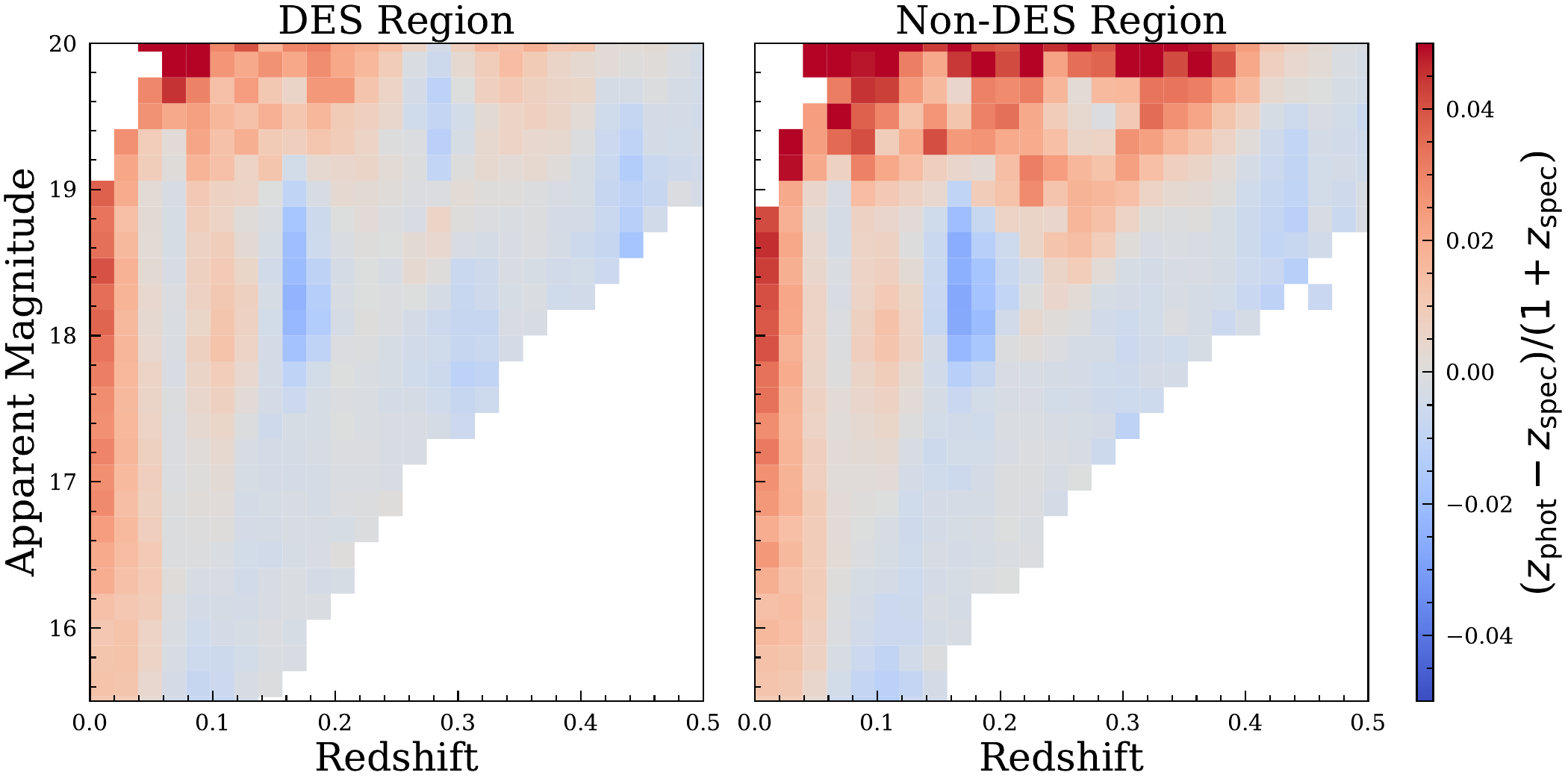}
    \caption{Redshift bias in $r$-band, as a function of apparent magnitude for \ac{DELVE}, separated by the \ac{des} (left) and non-\ac{des} (right) regions.}
    \label{fig:redshift_bias}
\end{figure}

In addition to measuring the redshift bias across all of \ac{DELVE}, we measure the redshift bias of the \ac{des} and non-\ac{des} regions of \ac{DELVE}, shown in Figure~\ref{fig:redshift_bias}. To compute the redshift bias, we use the same spectroscopic redshift catalogs from Section~\ref{sec:methods:catchoices:absmag}. We find minor differences in bias as a function of redshift and magnitude, based on the survey region considered, shown in Figure~\ref{fig:redshift_bias}.

To de-bias our redshift sample, we define the redshift bias $b$, where

\begin{align}
    b = \frac{z_{\text{photo}} - z_{\text{true}}}{1+z_{\text{true}}},\label{eq:pz_bias}
\end{align}

$z_{\text{photo}}$ is the photometric redshift of a given galaxy and $z_{\text{true}}$ is the true redshift of the source. For computing the redshift bias $b$, $z_{\text{true}}$ is taken as the spectroscopic redshift of the galaxy. 

After computing the median bias $b_{ij}$ as a function in apparent magnitude bin $m_i$ and true redshift bin $z_j$, we can apply a correction to the measured photometric redshift $z_{\text{photo}}$ by rearranging Equation~\ref{eq:pz_bias}:

\begin{align}
    z_{\text{true}} = \frac{z_{\text{photo},ij}-b_{ij}}{1+b_{ij}}
\end{align}

where $z_{\text{true}}$ is the de-biased redshift derived from the photometric redshift $z_{\text{photo},ij}$, and the bias correction $b_{ij}$. A different bias correction is applied to galaxies in the \ac{des} and non-\ac{des} region, based on the maps in Figure~\ref{fig:redshift_bias}.

\section{Results}
\label{sec:results}

\begin{table*}
    \centering
    \renewcommand{\arraystretch}{1.25} 
    \begin{tabular}{lccccc}
        \hline
        \hline
        Galaxy Catalog  & Photometric Band & Luminosity Function &Redshift Bias &  Dark & Dark + GW170817 \\
          & & Parameterization & Correction & $H_0$ [$\mathrm{km\,s^{-1}\,Mpc^{-1}}$] & $H_0$ [$\mathrm{km\,s^{-1}\,Mpc^{-1}}$] \\
        \hline
        Empty  & -- & -- & None & \HzDarkSpectralOfourb & \HzDarkBrightSpectralOfourb \\
        DES  & r-band & Static & None & \HzDarkDESStaticOfourb & \HzDarkBrightDESStaticOfourb \\
        DELVE & r-band & Static & None & \HzDarkDELVEstaticSchecterOfourb & \HzDarkBrightDELVEstaticSchecterOfourb \\
        DES  & r-band & Evolving & None & \HzDarkDESOfourb & \HzDarkBrightDESOfourb \\
        DELVE  & r-band & Evolving & None & \HzDarkDELVEOfourb & \HzDarkBrightDELVEOfourb \\
        DELVE & i-band & Evolving  & None & \HzDarkDELVEiBandOfourb & \HzDarkBrightDELVEiBandOfourb \\
        DELVE & r-band & Evolving  & De-biased by Region & \HzDarkDELVEDeBiased & \HzDarkBrightDELVEDeBiased \\
        \hline
    \end{tabular}
    \caption{$H_0$ results for different galaxy catalog choices for the empty-catalog condition, the DES catalog, and the DELVE catalog. Numerical constraints are median and 68\% CL of the posterior distribution.}
    \label{tab:fiducial_h0}
\end{table*}

For the galaxy catalog analysis, we adopt the choices listed in Table~\ref{tab:fiducial_h0}. For each configuration, we report the posterior on $H_0$ both from dark sirens alone and from dark sirens combined with the bright siren GW170817. 
GW170817 is excluded from the dark sirens only analysis.
The tabulated posteriors are reported in Table \ref{tab:fiducial_h0}.
All quoted $H_0$ values use the notation $H_0 = \mathrm{median}^{+84\rm th}_{-16\rm th}\,{\rm km\,s^{-1}\,Mpc^{-1}}$ ($1\sigma$-equivalent credible interval).

\subsection{Comparison to Previous Results}

For the empty galaxy catalog, we re-run the posterior instead of reusing the \ac{gwtc5} result. We find $H_0 = $ \HzDarkSpectralOfourb\ $\mathrm{km\,s^{-1}\,Mpc^{-1}}$, which is a slight difference relative to the reported \ac{gwtc5} posterior of $H_0=70.1^{+15.1}_{-13.3}$\ $\mathrm{km\,s^{-1}\,Mpc^{-1}}$. We note that the reported values from the \ac{LVK} \ac{gwtc5} analysis combines results from two pipelines, including \texttt{gwcosmo}.
The baseline determinations of $H_0$, the empty-catalog spectral-siren method, and the two static-luminosity-function analyses for \ac{des} and \ac{DELVE}, agree well with one another. The dark siren posterior for \ac{des} gives $H_0 = $ \HzDarkDESStaticOfourb\ $\mathrm{km\,s^{-1}\,Mpc^{-1}}$ compared to $H_0 = $ \HzDarkDELVEstaticSchecterOfourb\ $\mathrm{km\,s^{-1}\,Mpc^{-1}}$ for \ac{DELVE}. The three posteriors are consistent and none of the shifts are significant. We take this agreement as evidence that the choice of galaxy catalog, or its absence, is not driving the recovered value of $H_0$ in the static-luminosity-function case given the \ac{gwtc5} \ac{GW} sample.

\subsection{Evolving Luminosity Function Parameterization}

We next relax the assumption of a non-evolving galaxy luminosity function, parameterizing following Equations~\ref{eq:mstar_evolving}-\ref{eq:phistar_evolving} and using the parameters from Table~\ref{tab:lf_params}. 
The constraining power on $H_0$ decreases slightly for \ac{des} when changing from a static luminosity-function parameterization, moving from $H_0 = $ \HzDarkDESStaticOfourb\ $\mathrm{km\,s^{-1}\,Mpc^{-1}}$ when static to $H_0 = $ \HzDarkDESOfourb\ $\mathrm{km\,s^{-1}\,Mpc^{-1}}$ when evolving.
For \ac{DELVE}, moving from the static to evolving luminosity function parameterization shifts the dark-siren constraint from $H_0 = $ \HzDarkDELVEstaticSchecterOfourb\ $\mathrm{km\,s^{-1}\,Mpc^{-1}}$ to $H_0 = $ \HzDarkDELVEOfourb\ $\mathrm{km\,s^{-1}\,Mpc^{-1}}$, bringing it into closer agreement with the spectral-siren result, while modestly broadening the posterior.

\subsection{i-band Photometry}

Adopting the evolving luminosity function parameterization as our fiducial configuration, we evaluate the sensitivity to the photometric band used for the $K$-correction and absolute-magnitude weighting. To do this, we repeat the \ac{DELVE} evolving-luminosity-function analysis using $i$-band photometry in place of $r$-band. The luminosity function parameters for $i$-band are given in Table~\ref{tab:lf_params}. $i$-band provides a dark-siren result consistent with $r$-band: $H_0 =$ \HzDarkDELVEiBandOfourb\ $\mathrm{km\,s^{-1}\,Mpc^{-1}}$ ($i$-band) compared to $H_0 = $ \HzDarkDELVEOfourb\ $\mathrm{km\,s^{-1}\,Mpc^{-1}}$ ($r$-band). The two results are in close agreement, with the $r$-band posteriod slightly wider than the $i$-band case. We interpret this agreement as evidence that the measurement is not sensitive to the specific optical band chosen for the galaxy weighting, given the current \ac{GW} sample from \ac{gwtc5}.

\subsection{Redshift Bias Correction}

After applying a redshift bias correction, based on the features in magnitude-redshift space across the \ac{des} and non-\ac{des} regions, the constraints on $H_0$ show a moderate improvement. 
With the correction, we obtain $H_0= $ \HzDarkDELVEDeBiased\ $\mathrm{km\,s^{-1}\,Mpc^{-1}}$, broadly consistent with the uncorrected posterior, $H_0 = $ \HzDarkDELVEOfourb\ $\mathrm{km\,s^{-1}\,Mpc^{-1}}$. While consistent, the bias-corrected posterior constrains $H_0$ tighter than without the correction, and the resulting constraint is shifted to a lower value of $H_0$.



\section{Discussion}\label{sec:discussion}

\begin{figure*}
    \centering
    \includegraphics[width=\linewidth]{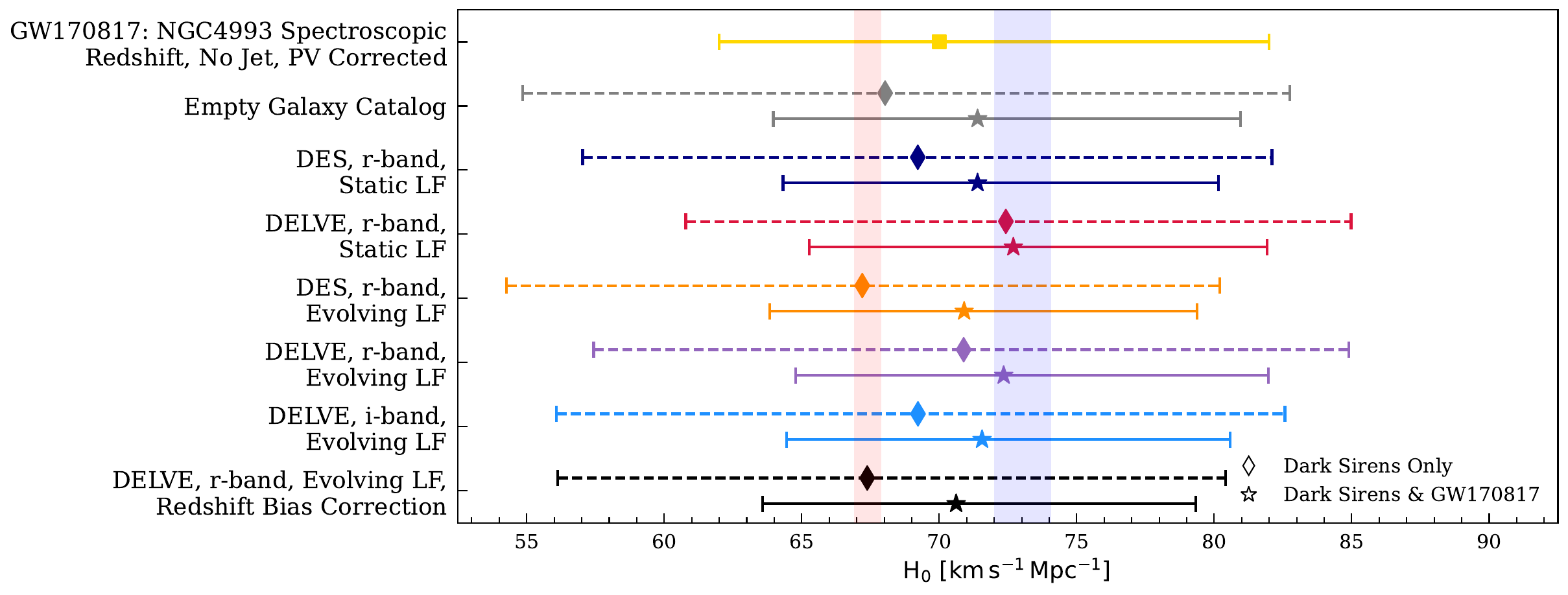}
    \caption{Constraints on $H_0$ from dark sirens (diamond) and dark sirens combined with GW170817 (star), shown for the different analysis configurations of DELVE, DES, and the empty catalog, which is equivalent to the spectral siren analysis. Tabulated results are reported in Table~\ref{tab:fiducial_h0}. The posterior of GW170817 is shown in a black square~\protect\citep{Abbott2017Nature}.  Planck and SH0ES constraints are shown for reference in the shaded bands~\protect\citep{Planck2020,Riess2022}. }
    \label{fig:h0_fiducial}
\end{figure*}

Across all seven analysis configurations, the recovered values of $H_0$ are consistent with one another, and varying the analysis choices within \ac{DELVE} does not significantly change the dark siren result. All of the posteriors, both dark-sirens alone and in combination with GW170817, overlap comfortably with both the \textit{Planck} (\citealt{Planck2020}) and SH0ES (\citealt{Riess2022}) constraints shown in Figure~\ref{fig:h0_fiducial}; the current statistical power is not yet sufficient to favor one constraint over the other.

Combining the dark-siren posteriors with the bright-siren measurement from GW170817 (black square, Figure~\ref{fig:h0_fiducial}) tightens the constraint substantially.
The empty-catalog combination yields $H_0 =$ \HzDarkBrightSpectralOfourb\ $\mathrm{km\,s^{-1}\,Mpc^{-1}}$. The constraint is improved when adding galaxy catalog information, with the tightest constraints produced by the \ac{des}, r-band, evolving luminosity function parametrization, with Chilingarian K-correction ($H_0=$ \HzDarkBrightDESOfourb\ $\mathrm{km\,s^{-1}\,Mpc^{-1}}$).
This convergence reflects the fact that GW170817's EM-counterpart-based posterior dominates the combined likelihood once it is included, and the high level of consistency across dark siren- results.

The robustness of the result to 
catalog choice (empty vs.\ \ac{des} vs.\ \ac{DELVE})
and photometric band ($r$ vs.\ $i$) is a useful validation: none of these choices, which affect how galaxy redshifts and absolute magnitudes are estimated, produces a shift larger than the statistical uncertainty on any individual measurement. As the catalog of \ac{GW} events grows to include more well-constrained mergers, we expect the catalog-dependent differences become more indicative of systematics.

The redshift bias correction, measured and applied based on survey region, accounts for general features based on the photometric quality across different survey regions of \ac{DELVE}.
The improved constraint after correcting for redshift bias reflects an improvement to the redshift precision, especially for bright galaxies, which carry a larger contribution to the in-catalog line of sight for luminosity-weighted galaxies, $\epsilon=1$.
Based on this result, we conclude that controlling and characterizing redshift bias is now a meaningful systematic for dark siren analyses. 



Given 
the measured evolution of the luminosity function, no significant change from using $i$-band photometry, the moderate improvement from a redshift bias correction, and that the culmination of these effects does not induce any systematic shift in the $H_0$ posterior,  we adopt 
$r$-band photometry and
an evolving luminosity function parameterization
with a redshift bias correction 
as our fiducial analysis configuration. 

The stability in constraining power from increasing the galaxy catalog area from \ac{des} to \ac{DELVE} could be explained by several factors. For one, the region of \ac{DELVE} that is not a part of \ac{des} has poorer quality photometry and photometric redshift precision. Additionally, the best localized dark siren, GW190814, lies completely within the \ac{des} footprint, meaning that other dark sirens that lie solely within the \ac{DELVE} footprint may introduce additional scatter in the $H_0$ posterior relative to the \ac{des}-only dark siren population. Including all data up to \ac{gwtc5}, \ac{DELVE} adds comprehensive coverage for only 3 additional \ac{GW} events from the top 10\% of well-localized mergers, shown in Figure~\ref{fig:delve_coverage}.


We directly measure an evolving luminosity function,  parameterize the evolution to be linear in redshift, and propagate this evolution into the redshift line-of-sight prior. An evolving luminosity function applied to \ac{DELVE} shifts the posterior only slightly compared to a static luminosity function under identical conditions, and the shift is well-within the statistical uncertainty. This suggests our results are not strongly sensitive to the luminosity function parameterization, though we retain the evolving parameterization as more physically motivated for a galaxy catalog spanning $z<0.5$, given the evolution we directly measure.


\section{Conclusion}\label{sec:conclusion}

We have extended the dark siren measurement of $H_0$ from the \ac{des} Y6 Gold catalog to the wider, photometrically heterogeneous \ac{DELVE} DR3 catalog, combined with \ac{GW} events from \ac{gwtc5}. Extending the dark siren method beyond the well-characterized \ac{des} footprint required formalizing three catalog-construction choices: a faint apparent-magnitude threshold of $m_{\rm thr,faint}=\mthr$, set directly by a requirement on the photometric-redshift bias of \ac{DELVE}, a bright apparent-magnitude threshold of $m_{\rm thr,bright}=\mthrbright$, set by the turnover in galaxy number counts due to CCD saturation, and a redshift bias correction based on survey region. With these choices, we obtain $H_0=$ \HzDarkFiducial\ $\mathrm{km\,s^{-1}\,Mpc^{-1}}$ from dark sirens alone, and $H_0 =$ \HzDarkBrightFiducial\ $\mathrm{km\,s^{-1}\,Mpc^{-1}}$ when combined with the bright siren GW170817. Across every catalog and configuration tested, the recovered $H_0$ posteriors are broad and mutually consistent.

The similarity between the empty-catalog and galaxy-catalog-informed posteriors reflects the small number of well-localized dark sirens currently available, and the underlying contribution of the spectral siren and bright siren constraints on the $H_0$ posterior. In this regime, the galaxy catalog is not yet a dominant contribution to the $H_0$ constraint, despite the measured improvement. This is also reflected in the analysis-configuration choices, where the differences between each analysis configuration show minor shifts for each scenario tested. This robustness is itself a useful validation of the method: none of the choices governing how individual galaxy redshifts and absolute magnitudes are produced results in a systematic shift in the $H_0$ posterior. All posteriors remain statistically consistent with constraints from the early- and late- universe, since the constraining power is not yet sufficient to discriminate between one class of constraints over the other.

Our faint-end threshold ($m_{\rm thr,faint}=$\mthr\ in $r$-band) is brighter than the faint limit used in the \ac{LVK} dark-siren analysis of GWTC-5.0 \citep{GWTC5_cosmo}, which uses a $10\sigma$ depth of $m_r=23.9$. Our choice on apparent magnitude limit is instead set directly by a requirement on photometric-redshift bias (Section~\ref{sec:methods:catchoices}), enabling us to safely extend the redshift range to $z=0.5$, matching the upper limit of the Chilingarian
$K$-correction.
On the bright end, we impose an explicit, empirically derived threshold, $m_{\rm thr,bright}=$\mthrbright, identified from the turnover in the $r$ and $i$-band galaxy number counts (Figure~\ref{fig:mag_cuts}). Previous analyses use fixed apparent magnitude thresholds and $z_{low}>0$ to exclude saturated sources at low redshift. By imposing a bright apparent magnitude threshold that alters the in and out-of-catalog term of the redshift line-of-sight prior, we include galaxy catalog information below $z=0.05$ with deep photometric galaxy catalogs for the first time, providing additional cosmological information for nearby \ac{GW} events that carry a larger fraction of the constraining power on $H_0$ through their small localization volumes.
Additionally, we find that redshift bias plays an increasingly large role in the preparation of galaxy catalogs for dark siren cosmology. The bias-correction in this work corrects for many global features of the redshift distribution, but could be augmented in future work by higher quality photometry across a survey, or homogeneous imaging data that does not require treatment for spatially dependent redshift quality.


This work demonstrates that the dark siren method can be extended from a single, deep, and well-characterized survey footprint like \ac{des} to a larger and less photometrically uniform galaxy catalog like \ac{DELVE} without a loss of robustness. This extension is possible only if catalog construction choices that govern redshift precision and spurious object contamination are closely controlled.

As the number of well-localized \ac{GW} events grows over future observing runs, we expect galaxy-catalog choices to become correspondingly more informative. Along with informativeness, these catalog-level choices are likely to reveal systematics, in contrast to the current level of precision. The catalog construction criteria applied to \ac{DELVE} act as a template, directly applicable to forthcoming, wider and deeper photometric surveys, including the \textit{\ac{lsst}}~\citep{Ivezic2019LSST}, \textit{Euclid}~\citep{2022A&A...662A.112E}, and \textit{Roman}~\citep{2025arXiv250510574O}. Extending the dark siren method to future galaxy surveys, combined with more and better-localized \ac{GW} events from future \ac{LVK} observing runs, should enable galaxy catalog information to become a more prevalent contribution to cosmological constraints, sharpening stardard siren measurements and providing an additional measurement for comparison to early- and late-universe measurements.
\begin{acknowledgements}
\begingroup\let\section\aaDropStarSection
\section*{Acknowledgements}

We thank the DELVE collaboration, especially Chin-Yi Tan and Alex Drlica-Wagner for their consultation and perspectives on the DELVE DR3 data products essential to this analysis. We thank the LVK Cosmology Working Group for the productive discussions throughout the development of this work. 

Funding for this work was provided by the University of Zurich (UZH) and the Swiss National Science Foundation (SNSF) under grant number 10002981. DL was supported by the UZH Postdoc Grant, grant number K-72341-01-01. MB is supported by the Polish National Science Center through grant no. 2020/38/E/ST9/00395 and by the Polish Ministry of Science and Higher Education under the agreement no. 2026/WK/06. S.F. received support from the French government under the France 2030 investment plan, as part of the Excellence Initiative of Aix Marseille University - amidex (AMX-19-IET-008 - IPhU).
The work of GD was supported by the PHC BALATON 2025 (52886UM) and 2024-1.2.5-TÉT grants. GD acknowledges the OSCARS project, which has received funding from the European Commission’s Horizon Europe Research and Innovation programme under grant agreement No. 101129751.

This project used public archival data from the Dark Energy Survey (DES) and the DECam Local Volume Exploration Survey (DELVE). Funding for the DES Projects has been provided by the U.S. Department of Energy, the U.S. National Science Foundation, the Ministry of Science and Education of Spain, the Science and Technology Facilities Council of the United Kingdom, the Higher Education Funding Council for England, the National Center for Supercomputing Applications at the University of Illinois at Urbana-Champaign, the Kavli Institute for Cosmological Physics at the University of Chicago, the Center for Cosmology and Astro-Particle Physics at the Ohio State University, the Mitchell Institute for Fundamental Physics and Astronomy at Texas A\&M University, Financiadora de Estudos e Projetos, Funda\c{c}\~ao Carlos Chagas Filho de Amparo \`a Pesquisa do Estado do Rio de Janeiro, Conselho Nacional de Desenvolvimento Cient\'ifico e Tecnol\'ogico and the Minist\'erio da Ci\^encia, Tecnologia e Inova\c{c}\~ao, the Deutsche Forschungsgemeinschaft, and the collaborating institutions in the Dark Energy Survey.

This material is based upon work supported by NSF’s LIGO Laboratory, which is a major facility fully funded by the National Science Foundation. Work herein is based in part on observations at Cerro Tololo Inter-American Observatory, NOIRLAB, which is operated by the Association of Universities for Research in Astronomy (AURA) under a cooperative agreement with the NSF.

This work made use of infrastructure services provided by the UZH Science IT team (\href{www.s3it.uzh.ch}{www.s3it.uzh.ch}).

\endgroup
\end{acknowledgements}
\section*{Data Availability}

The DELVE DR3 galaxy catalog used in this analysis is publicly available via the NOIRLab Astro Data Lab\footnote{\href{https://datalab.noirlab.edu/data/delve}{https://datalab.noirlab.edu/data/delve}}; the associated data-release publication was not yet available at the time of writing Section~\ref{sec:data:delve}), though the general DR3 properties are described in \citet{Tan2025}. The DES Y6 Gold catalogue is described in \citet{Bechtol2026DESY6Gold}, and available for download from CosmoHub\footnote{\href{https://cosmohub.pic.es/catalogs/333}{https://cosmohub.pic.es/catalogs/333}}. 
GW parameter estimation samples can be downloaded from Zenodo for GWTC-5.0\footnote{\href{https://zenodo.org/records/20276106}{https://zenodo.org/records/20276106}, \href{https://zenodo.org/records/20348006}{https://zenodo.org/records/20348006}}, GWTC-4.0\footnote{\href{https://zenodo.org/records/17014085}{https://zenodo.org/records/17014085}}, GWTC-3.0\footnote{\href{https://zenodo.org/records/8177023}{https://zenodo.org/records/8177023}}, and GWTC-2.1\footnote{\href{https://zenodo.org/records/6513631}{https://zenodo.org/records/6513631}} as well as the injections used to model the GW selection effects\footnote{\href{https://zenodo.org/records/19500052}{https://zenodo.org/records/19500052}}.
The spectroscopic redshift catalogs used to assess photometric redshift bias are publicly available from their respective survey data releases. 
This work makes use of \texttt{gwcosmo} \citep{Gray2022,Papadopoulos2026}.
Scripts used for catalog preparation, luminosity-function fitting, line-of-sight prior construction, and posterior sampling can be shared upon reasonable request to the corresponding author.

\bibliographystyle{aa}
\bibliography{biblio}

\appendix

\end{document}